\documentclass[runningheads]{llncs}
\usepackage[T1]{fontenc}
\usepackage{graphicx,verbatim}
\usepackage{hyperref} 
\usepackage{color}

\usepackage[table]{xcolor}
\usepackage{amsmath}
\usepackage{cases}
\usepackage{multirow}
\usepackage{tabularray}
\usepackage{xcolor}
\usepackage{svg}
\usepackage{bm}
\usepackage{amssymb}
\usepackage{booktabs}
\usepackage{bbding}

\begin{document}
%
\title{ConStyX: Content Style Augmentation for Generalizable Medical Image Segmentation}
\titlerunning{Content Style Augmentation}
%

\author{Xi Chen\inst{1,2,\dag} \and
Zhiqiang Shen\inst{1,2,\dag} \and
Peng Cao\inst{1,2,3(}\Envelope\inst{)} \and
Jinzhu Yang\inst{1,2,3} \and
Osmar R. Zaiane\inst{4}}
\footnotetext[1]{$\dagger$ Xi Chen and Zhiqiang Shen contributed equally to this work.}

\authorrunning{X. Chen et al.}
%
\institute{School of Computer Science and Engineering, Northeastern University, Shenyang, China \and
Key Laboratory of Intelligent Computing in Medical Image of Ministry of Education, Northeastern University, Shenyang, China \and
National Frontiers Science Center for Industrial Intelligence and Systems Optimization, Shenyang, China \\
\email{caopeng@mail.neu.edu.cn} \and
Alberta Machine Intelligence Institute, University of Alberta, Edmonton, Canada}

\maketitle              
\begin{abstract}
Medical images are usually collected from multiple domains, leading to domain shifts that impair the performance of medical image segmentation models. Domain Generalization (DG) aims to address this issue by training a robust model with strong generalizability. Recently, numerous domain randomization-based DG methods have been proposed. However, these methods suffer from the following limitations: 
1) constrained efficiency of domain randomization due to their exclusive dependence on image style perturbation, and 2) neglect of the adverse effects of over-augmented images on model training. 
To address these issues, we propose a novel domain randomization-based DG method, called content style augmentation (ConStyX), for generalizable medical image segmentation. 
Specifically, ConStyX 1) augments the content and style of training data, allowing the augmented training data to better cover a wider range of data domains, and 2) leverages well-augmented features while mitigating the negative effects of over-augmented features during model training. 
Extensive experiments across multiple domains demonstrate that our ConStyX achieves superior generalization performance. The code is available at \href{https://github.com/jwxsp1/ConStyX}{\textit{\texttt{https://github.com/jwxsp1/ConStyX}}}.

\keywords{Domain generalization  \and Domain randomization \and Medical image segmentation \and Deep features.}

\end{abstract}
\section{Introduction}
Medical image segmentation is an important task in computer-aided diagnosis and treatment. In recent years, this field has witnessed significant advancements, attributed to the progress of deep learning \cite{asgari2021deep}. However, learned segmentation models encounter significant performance drops when the training and test sets are sampled from different distributions, where domain shifts arise due to variations in acquisition processes and patient populations \cite{guan2021domain,xie2021survey}. Domain generalization (DG) has been proposed to improve the generalizability of segmentation models, with the setting that a model is trained using data from single or multiple domains and tested on unseen domains \cite{zhou2022domain,wang2022generalizing}.
\begin{figure}[ht]
    \centering                                    
    \includegraphics[width=\linewidth]{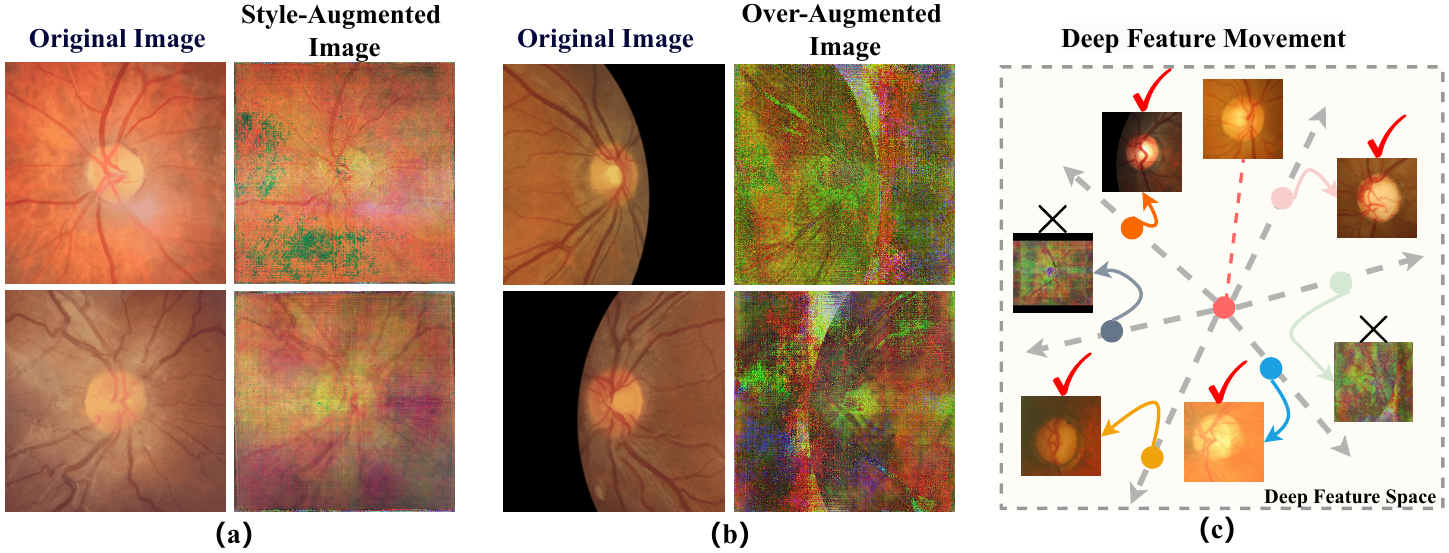}    
    \caption{Visualization of original and augmented images. (a) Original images (left) and style-augmented images (right). (b) Original images (left) and over-augmented images (right). (c) In the deep feature space, when the feature vector of a sample moves along a specific direction (gray dashed arrow), the resulting feature vector corresponds to an augmented image ({\color{red}$\checkmark$}: well-augmented images or $\times$: over-augmented images).}\label{fig:figure1} 
\end{figure}

Existing DG methods can be categorized into decoupling-based \cite{choi2021robustnet,peng2022semantic,hu2023devil} and domain randomization-based methods \cite{zhao2024morestyle,zhou2021domain,jia2025dginstyle}. 
Decoupling-based methods attempt to extract domain-invariant features from data by normalizing the features or constructing dedicated modules within the model. However, these approaches may compromise the semantic information within the features, as there is no guarantee that only domain-specific features are eliminated, leading to a degradation in the model's discriminative capability. 
In contrast, domain randomization-based DG methods aim to simulate unseen domains using source domain data. This line of approaches can be divided into two branches: 1) image perturbation via image reconstruction \cite{chen2022maxstyle,zhao2024morestyle} or generation \cite{jia2025dginstyle}, and 2) feature perturbation by transferring statistical information \cite{zhou2021domain,li2022uncertainty,zhang2022exact,zhang2024domain}. 
However, these methods only augment image style [Fig.~\ref{fig:figure1}(a)], limiting the diversity of augmented data distributions.
Moreover, due to the uncontrollability of the perturbation process, it is possible to produce over-augmented images with corrupted semantic information and unreal image appearance [Fig.~\ref{fig:figure1}(b)], which degrade the model training. 

To address these issues, we propose a novel content style augmentation method (ConStyX) for generalizable medical image segmentation. 
Our method is built upon the following assumption: moving a deep feature along a certain direction produces a new feature that corresponds to another sample of the same class but contains different content\footnote{"Content" refers to substructures within a specific class region. For example, different optic discs (class) may contain distinct optic vessels (content) in fundus images.} and style information \cite{upchurch2017deep,wang2021regularizing} [Fig.~\ref{fig:figure1}(c)]. 
Specifically, ConStyX consists of two components: 
1) Deep Feature Augmentation algorithm (DFA) conducts content and style augmentation, ensuring the augmented data covers a wide scope for unseen target domains and 
2) Augmented Feature Utilization strategy (AFU) quantifies the contributions of augmented features to the model training, thereby mitigating the negative impacts of over-augmented features while fully leveraging the well-augmented ones. 
We evaluated our method on five public fundus datasets, corresponding to five different domains. Extensive experiments demonstrate that our method achieves better generalization performance on unseen target domains compared with state-of-the-art domain generalization methods.

Our contributions are three-fold:
\begin{itemize}  
    \item We propose a content style augmentation method that generates augmented deep features by moving the original deep features toward correct directions with appropriate degrees, thus enabling the training data to cover a wider range of unseen domains.  
    \item We devise an augmented feature utilization strategy to quantify the contributions of augmented features to model training, for exploiting well-augmented features while mitigating the negative effects of over-augmented ones.  
    \item Our ConStyX outperforms the baseline and five state-of-the-art DG methods on the joint optic disc (OD) and optic cup (OC) segmentation benchmarks.
\end{itemize}  

\section{Method}
\textbf{{Notions \& Notations.}} Given a source domain $D=\{(X_i, Y_i)^M_{i=1}\}$, for single domain generalized medical segmentation, the objective is to learn a segmentation model $f(\cdot;\theta)$ from  $D$ that shows strong generalization capability on unseen domains. The segmentation model $f(\cdot;\theta)$ consists of an encoder $E(\cdot;\theta_1)$ and a segmentation head $H(\cdot;\theta_2)$. Let $Z_i \in R^{N\times H\times W}$ be the deep features, where $N$, $H$, and $W$ denote the channel, height, and width respectively, and $\textbf{z}_i^j \in R^{N\times 1 \times 1}$ be the deep feature of the $j^{th}$ pixel for $X_i$ with class $c$.

\begin{figure}[ht]
    \centering                                    
    \includegraphics[width=\linewidth]{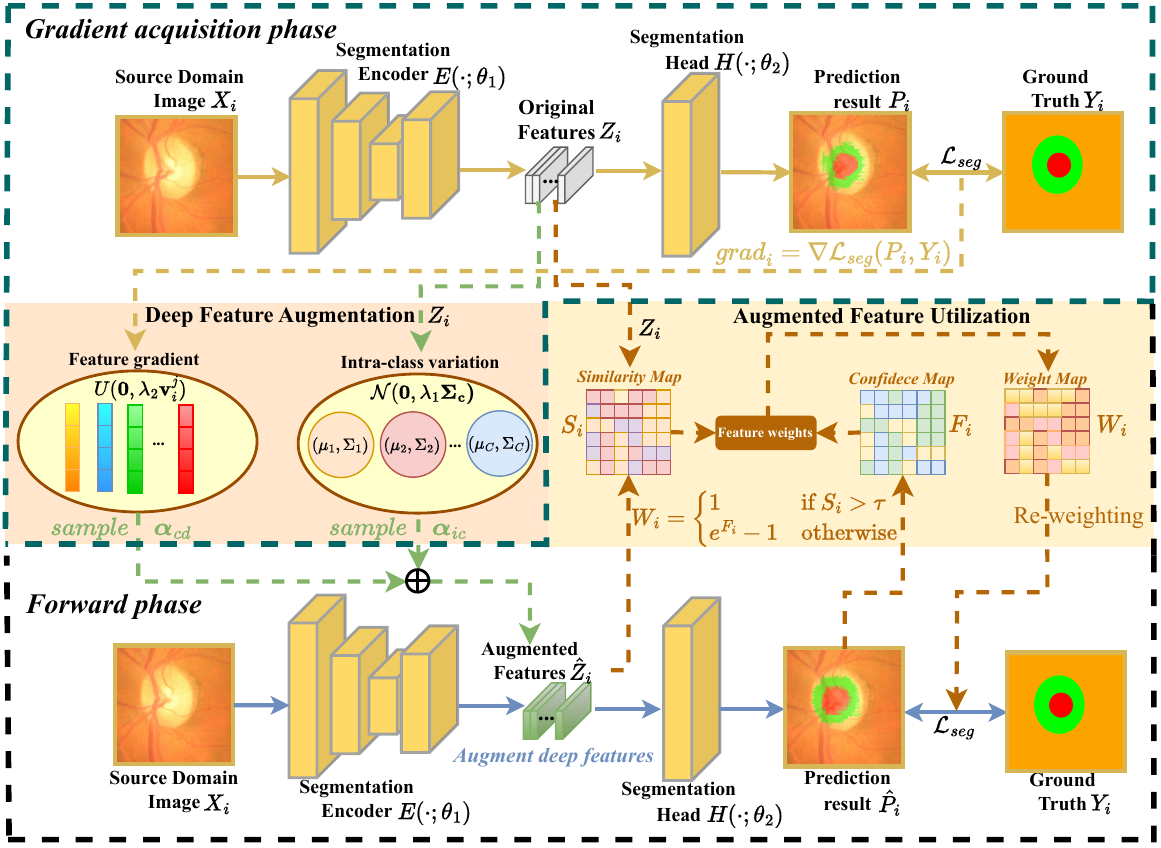}    
    \caption{Overview of the proposed Content Style Augmentation framework (ConStyX). It includes: 1) Deep Feature Augmentation algorithm (DFA) for content-style augmentation under the guidance of intra-class variation and feature gradient and 2) Augmented Feature Utilization strategy (AFU) for augmented feature re-weighting.} \label{fig:overview} 
\end{figure}

\noindent
\textbf{{Overview.}} Considering the limited diversity in style augmentation and the detrimental effects of over-augmented samples, this work aims to develop an effective and controllable content-style augmentation method for generalizable medical image segmentation. 
Our core assumption is that: moving a deep feature along a specific direction generates a new feature corresponding to an augmented image of the same class but with distinct style and content characteristics.
Based on this, we propose a content and style augmentation framework (ConStyX). 
As illustrated in Fig.~\ref{fig:overview}, ConStyX consists of two key components: 1) Deep Feature Augmentation algorithm (DFA) for intra-domain and cross-domain feature movements under the guidance of intra-class variation and feature gradient and 2) Augmented Feature Utilization strategy (AFU) for augmented feature re-weighting according to their contributions to model training quantified by feature similarity and prediction confidence.

\subsection{Deep Feature Augmentation (DFA)}
The fundamental concept of DFA lies in 
\textit{How to appropriately move the deep features while preserving their original semantic information.} 
To this end, we determine the direction and degree of movement by \textbf{intra-class variation} and \textbf{feature gradient}. 
Specifically, we construct two augmentation distributions, i.e., 
 $\mathcal{N}(\mathbf{0},\lambda_1 \bm{\Sigma}_{c})$ and $U(\mathbf{0},\lambda_2 \textbf{v}_i^j)$, from which movement vectors can be sampled for feature augmentation: 
\begin{align}
\hat{\textbf{z}}_i^j=\mathbf{z}_i^j +\bm{\alpha}_{ic}+\bm{\alpha}_{cd}, \quad \bm{\alpha}_{ic} \sim \mathcal{N}(\mathbf{0},\lambda_1 \bm{\Sigma}_{c  }), \quad \bm{\alpha}_{cd} \sim U(\mathbf{0},\lambda_2 \mathbf{v}_i^j) \label{eq:eq7}
\end{align}
where $\bm{\alpha}_{ic}$ represents the intra-domain augmentation vector determined by capturing the maximum \textbf{intra-domain variation}, 
and $\bm{\alpha}_{cd}$ denotes the cross-domain augmentation factor guided by \textbf{feature gradients}, $\bm{\Sigma}_{c}$ refers to a class-conditional covariance, $\mathbf{v}_i^j$ indicates feature mask, and $\lambda_1$ and $\lambda_2$ are two scaled factors.
Through these movements, deep features are appropriately augmented along the correct directions with proper degrees, generating augmented features that correspond to content-style augmented samples in the image space.

\paragraph{\textbf{Intra-class variation.}} 
It yields the movement vector $\bm{\alpha}_{ic} \sim \mathcal{N}(\mathbf{0},\lambda_1 \bm{\Sigma}_{c  })$ to enable features to move along the maximum intra-class variation direction.
Concretely, we extract a feature map $Z_i$ from image $X_i$ and establish a zero-mean multivariate normal distribution for the feature points corresponding to a certain class based on the label $Y_i$. 
Let $\bm{\mu}^{(t)}_{c}$ and $\bm{\Sigma}^{(t)}_{c}$ represent the mean and covariance matrix of the ${c}^{th}$ class features in the first $t$ iterations, which are estimated using online weighted averaging; $\bar{\bm{\mu}}^{(t)}_{c}$ and $\bar{\bm{\Sigma}}^{(t)}_c$ are the mean and covariance matrix of the features of ${c}^{th}$ class in the $t^{th}$ iteration;
$n^{(t)}_{c}$ denotes the total number of deep features belonging to ${c}^{th}$ class in the first $t$ iterations, and $m^{(t)}_{c}$ denotes the number of deep features belonging to ${c}^{th}$ class in the $t^{th}$ iteration. For the $c^{th}$ class feature points, we establish the multivariate normal distribution $\mathcal{N}(\mathbf{0},\lambda_1\bm{\Sigma}_{c})$, where the class-conditional covariance $\bm{\Sigma}_{c}$ in the first $t$ iterations is obtained by:
\begin{align}
\bm{\Sigma}^{(t)}_{c}=\frac{n^{(t-1)}_{c  }\bm{\Sigma}^{(t-1)}_{c  }+ m^{(t)}_{c  } \bar{\bm{\Sigma}}^{(t)}_{c  }}   
{n^{(t-1)}_{c  }+m^{(t)}_{c  }}+
              \frac{n^{(t-1)}_{c  } m^{(t)}_{c  }\Delta\bm{\mu}_{c  }\Delta\bm{\mu}_{c  }^\top}{(n^{(t-1)}_{c  }+m^{(t)}_{c  })^2}
\end{align}
where $\bm{\mu}^{(t)}_{c}=\frac{n^{(t-1)}_{c}\bm{\mu}^{(t-1)}_{c}+ m^{(t)}_{c} \bar{\bm{\mu}}^{(t)}_{c}}{n^{(t-1)}_{c}+m^{(t)}_{c}}$, 
$\Delta\bm{\mu}_{c}=\bm{\mu}^{(t-1)}_{c}- \bar{\bm{\mu}}^{(t)}_{c}$,
and $n^{(t)}_{c}=n^{(t-1)}_{c}+m^{(t)}_{c}$.
\paragraph{\textbf{Feature gradient.}} 
By capturing the intra-class variations, the features can be forced to move along the direction of the most significant intra-class variation. However, this strategy limits the augmentation to be operated within the source domain.
To further expand the augmented training data distribution over unseen domains, we introduce gradient-guided feature movement to generate another moving vector $\bm{\alpha}_{cd} \sim U(\mathbf{0},\lambda_2 \mathbf{v}_i^j)$, which is applied along the minimum gradient directions to ensure cross-domain and semantic-invariant content style augmentation.
Specifically, we first obtain the gradient of $\textbf{z}_i^j$: ${grad}_i^j=\nabla \mathcal{L}_{seg}(P_i^j,Y_i^j)$, where $\mathcal{L}_{seg}$ denote the segmentation loss and $P_i^j$ represents the model prediction. 
Note that model parameters are not updated in this process. 
Then, we define the feature mask $\mathbf{v}_i^j\in R^{N\times1\times1}$ as:
\begin{align}
{(\mathbf{v}_i^j)}_n=\begin{cases} 
  1 & \text{if} \,\, n\in {pos}_i^j \\
  0 &  \text{otherwise} 
\end{cases}
\end{align}
where the position of the minimum $k$ partial derivatives in $\mathbf{z}_i^j $ are obtained by: ${pos}_i^j ={\rm MinSort}(grad_i^j)[:k]$. ${\rm MinSort}(\cdot)$ is an ascending sorting function. 

Based on the feature mask $\mathbf{v}_i^j$, we construct the uniform distribution $U \sim (\textbf{0},\lambda_2 \textbf{v}_i^j)$ to for further feature augmentation, where $\lambda_2$ denotes scaled factor. 


\subsection{Augmented Feature Utilization (AFU)}
Although DFA aims to move deep features in appropriate directions for content style augmentation, it inevitably generates new features with distinct characteristics. 
We categorize these features into three types: 1) \textbf{trivial-augmented features}: maintain similar information with the original features,  2) \textbf{over-augmented features} lose the original information and even become noise, and 3) \textbf{well-augmented features} augment the content and style information appropriately while maintaining their original semantic information.
It is crucial to suppress the adverse effects of over-augmented features during model training while sufficiently exploiting the well-augmented ones. 

To this end, we introduce AFU, which leverages both \textbf{feature similarity} and \textbf{prediction confidence} to determine the contributions of the three types of augmented features to model training. 
Specifically, we compute the \textbf{cosine similarity} map $S_i$ between an original deep feature map $Z_i$ and its augmented counterpart $\hat{Z}_i$, as well as the \textbf{prediction confidence} ${F}_i=1-{\rm MinMax}(-\sum_{m=1}^C (\hat{P}_i)_m log((\hat{P}_i)_m))$ for augmented feature $\hat{Z}_i$ (where ${\rm MinMax(\cdot)}$ denotes a Min-Max normalization operation, and ${F}_i$ is normalized to $[0, 1]$). Formally, based on $S_i$ and $F_i$, the three type augmented features are defined as:
1) the augmented feature points with a cosine similarity higher than threshold $\tau$ are considered trivial-augmented features,
2) those with a similarity lower than $\tau$ and larger prediction confidence are divided as well-augmented features, 
and 3) those with a similarity lower than $\tau$ and lower confidence are regarded as over-augmented features. 
During the training process, the weight $W_i$ is employed to determine the contributions of the augmented feature $\hat{Z}_i$ to model training:
\begin{align}
W_i^j=\begin{cases} 
  1 & \text{if} \,\, {S}_i^j > \tau \label{eq:eq8} \\
  e^{{F}_i^j}-1 &  \rm otherwise 
\end{cases}
\end{align}
which serves as a pixel-wise weight for the segmentation loss (combining cross-entropy and Dice loss).

\section{Experiment}
\textbf{Datasets and Evaluation Metrics.}
\textit{Datesets: } We use five fundus datasets \cite{REFUGE,retinal,ORIGA,Drishti-GS} corresponding to five different domains for joint segmentation of optic cup (OD) and optic disc (OC): BinRushed (195 images), Magrabia (95 images), REFUGE (400 images), ORIGA (650 images), and Drishti-GS (101 images); each image is resized to $512\times512$.
We adopt the extremely challenging single-domain generalization setting, where one of the datasets is considered the source domain and divided into training and validation sets with a 9:1 ratio, while the remaining datasets serve as a test set.
\textit{Evaluation metrics: } Dice similarity coefficient (DSC, \%).

\noindent
\textbf{Implementation Details.}
\textit{Experimental environment:} All experiments are conducted using PyTorch \cite{paszke2019pytorch} on a Tesla V100 with 32GB GPU memory. U-Net \cite{ronneberger2015u} with a modified ResNet-34 encoder \cite{he2016deep} is used as the segmentation backbone for both our ConStyX and all compared methods.
\textit{Hyperparameter setting:} We set $k = 5$, $\lambda_1 = 1$, and $ \lambda_2 = 0.5$ for the DFA module, and the threshold $\tau = 0.6$ for the AFU module. 
ConStyX is optimized using the SGD optimizer with a momentum of 0.99 and an initial learning rate of 0.001 decayed according to a polynomial rule. 
The batch size and the number of training epochs are 8 and 100, respectively.

\subsection{Comparison with other DG methods}
We conduct comparative experiments over various state-of-the-arts, including: statistics transfer-based (MixStyle \cite{zhou2021domain}, DSU \cite{li2022uncertainty}, EFDM \cite{zhang2022exact}, TriD \cite{chen2023treasure}, and CSU \cite{zhang2024domain}),  random convolution-based (RandConv \cite{xu2020robust}), adversarial noise-based (MoreStyle \cite{zhao2024morestyle}), and feature disentanglement-based (CCSDG \cite{hu2023devil}) methods.

\begin{figure}[!t]
    \centering                                    
    \includegraphics[width=\linewidth]{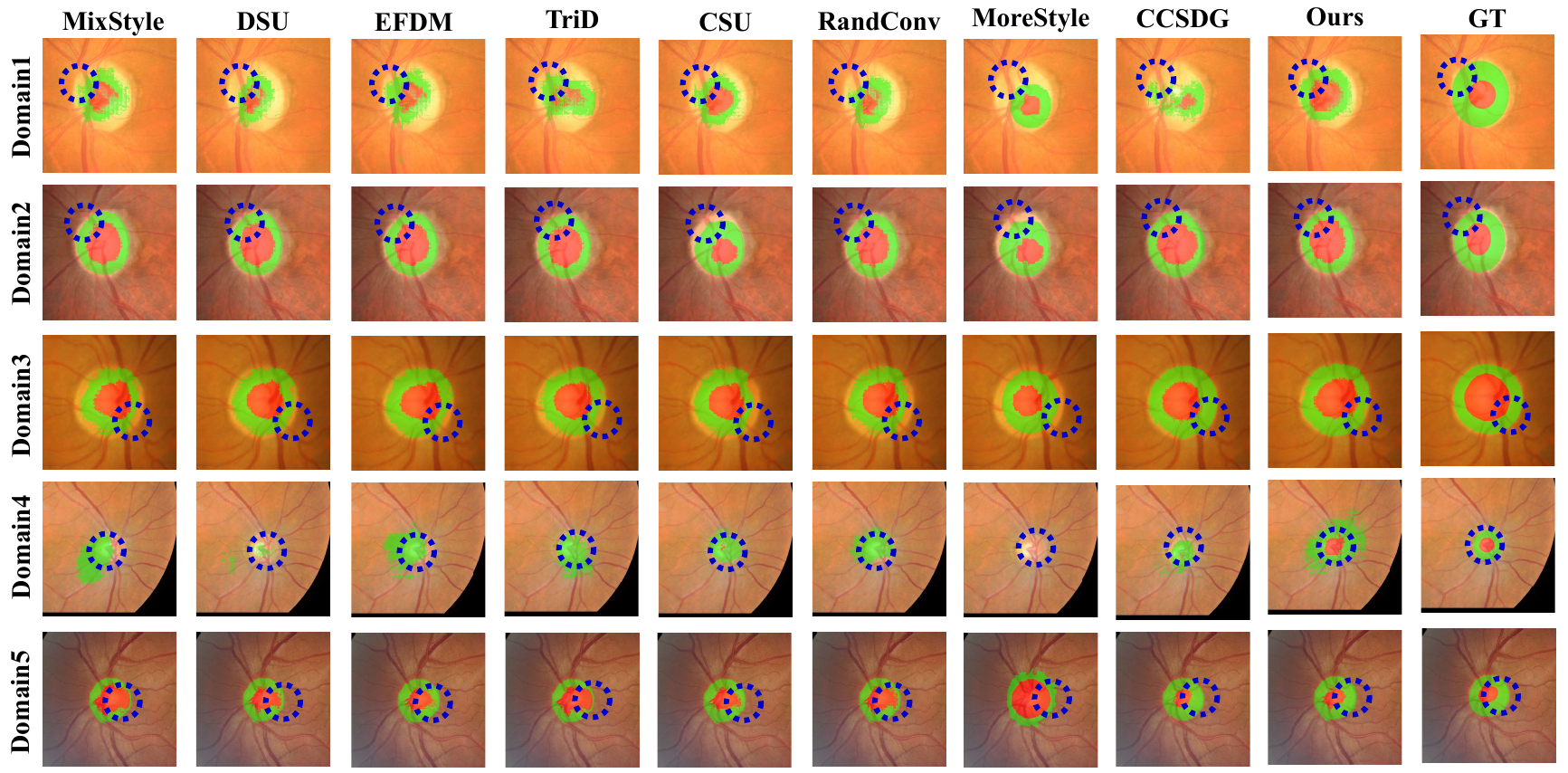}    
    \caption{Visualization of OD and OC segmentation results. Blue dashed circles highlight some regions with significant differences in the segmentation results.} 
    \label{fig:results} 
\end{figure}

Overall, both the segmentation performance in Table~\ref{tab:table1} and the qualitative results in Fig.~\ref{fig:results} suggest that our ConStyX consistently outperforms other DG methods, showing its superior generalization capability for cross-domain medical image segmentation. 
Specifically, the five statistical information transferred-based methods and the random convolution-based method yield similar segmentation results. 
This phenomenon is attributed to their limited augmentation efficacy (only augmenting image style), resulting in augmented samples that only cover a narrow range of unseen domains.
Due to the adversarial noise generating augmented samples that deviate significantly from real images, MoreStyle achieved unsatisfactory results. Meanwhile, CCSDG attains the second-highest average DSC across domains, i.e., 76.19 \%, by integrating feature disentanglement and various style augmentation techniques. 
In contrast, our ConStyX outperforms the second-highest result by 2.09\% in terms of DSC, due to the advantages of the content-style augmentation to generate diverse augmented samples and the feature re-weighting strategy to fully leverage well-augmented features.

\begin{table}[!t]
\centering
\caption{Comparative results of cross-domain segmentation performance (OD, OC). 
The best results are highlighted in \textbf{bold}.
}
\resizebox{\linewidth}{!}{
    \begin{tabular}{c| c c c c c |c}
    \toprule[1pt]
    \multirow{2}*{Method}&Domain1& Domain2& Domain3& Domain4& Domain5& Average\\
    \cline{2-7}
     &DSC $\uparrow$& DSC $\uparrow$& DSC $\uparrow$& DSC $\uparrow$& DSC $\uparrow$& DSC $\uparrow$ \\
     \midrule[1pt]
     MixStyle \cite{zhou2021domain}&(86.67, 64.86) & (87.34, 73.78)
& (86.83, 67.34) & (78.03, 59.49) & (83.52, 67.27)& 75.51\\
     DSU \cite{li2022uncertainty}&(86.64, 65.99) & (87.44, 74.20)
& (86.38, 66.80)& (78.77, 57.98)& (83.44, 66.00)& 75.36\\
    EFDM \cite{zhang2022exact}&(86.32, 65.11) &(88.03, 75.62)
& (86.55, 67.34)& (77.02, 58.22)& (83.66, 67.60)& 75.55\\
    TriD \cite{chen2023treasure}&(84.03, 63.73) & (85.75, 70.16)
& (86.93, \textbf{67.70})& (75.46, 66.96)& (84.85, 57.38)&74.30\\ 
CSU \cite{zhang2024domain}&(87.58, 69.19) & (87.50, 74.91)
& (86.87, 67.55)& (79.58, 57.93)& (83.50, 66.46)& 76.11 \\ 
RandConv \cite{xu2020robust}&(85.87, 65.99) & (88.10, 76.05)
& (86.47, 65.77)& (79.33, 57.79)& (83.30, 67.52)& 75.62\\
    MoreStyle\cite{zhao2024morestyle}&(80.38, 59.60) & (88.47, 64.32)
& (82.63, 63.85)& (77.07, 51.91)& (78.14, 51.63)& 69.80\\  
 CCSDG \cite{hu2023devil}&(86.21, 63.55) & (\textbf{90.34}, \textbf{78.24})
& (87.18, 65.44)& (81.00, 63.16)& (82.21, 64.57)&76.19 \\ \midrule
   ConStyX (ours) &({\textbf{88.95}, \textbf{72.55}})& ({89.86}, {77.61}) & (\textbf{88.17}, 67.22)& (\textbf{81.09}, \textbf{67.50})& (\textbf{86.67}, \textbf{69.19})&\textbf{78.88}\\ 
    \bottomrule[1pt]
    \end{tabular}
    }
\label{tab:table1}
\end{table}

\subsection{Ablation study}
\noindent \textbf{Analysis of the proposed components.} We conduct an ablation study on the five domains to evaluate the effect of our proposed modules including DFA and AFU. In Table~\ref{tab:table2}, one can observe that the segmentation performance gradually increases as each component is incorporated into our method. 
Compared with the baseline, our final model obtains average segmentation improvements of 9.17\% and 9.78\% across five distinct domains, significantly improving the generalizability of the baseline model.
\begin{table}[!t]
\centering
\caption{Performance (OD, OC) of ablation study on joint segmentation of OD and OC. The best results are highlighted in \textbf{bold}.}
\resizebox{\linewidth}{!}{
    \begin{tabular}{c| c c c c c |c}
    \toprule[1pt]
    \multirow{2}*{Method}&Domain1& Domain2& Domain3& Domain4& Domain5& Average\\
    \cline{2-7}
     &DSC $\uparrow$& DSC $\uparrow$& DSC $\uparrow$& DSC $\uparrow$& DSC $\uparrow$& DSC $\uparrow$ \\
     \midrule[1pt]
     Baseline&(86.21, 53.02)& (85.81, 58.05)& (78.73, 50.80) 
& (78.41, 57.56) & (81.72, 60.72) & 69.10 \\
    Baseline+DFA &(87.87, 69.39) & (\textbf{89.90}, 76.85)
& (88.11, 66.89)& (79.98, \textbf{68.03})& (\textbf{86.67}, 69.04)& 78.27 \\ 
   Baseline+DFA+AFU&({\textbf{88.95}, \textbf{72.55}})& ({89.86}, \textbf{77.61}) & (\textbf{88.17}, \textbf{67.22})& (\textbf{81.09}, {67.50})& (\textbf{86.67}, \textbf{69.19})&\textbf{78.88 }\\ 
    \bottomrule[1pt]
    \end{tabular}
    }
\label{tab:table2}
\end{table}

\noindent \textbf{Investigation of distribution forms.} 
To investigate the influence of distribution forms on feature movement operations, we conduct a comparative experiment on the feature gradient-guided movement by constructing two gradient-guided augmentation distributions: a uniform distribution $\bm{\alpha}_{cd} \sim U(\mathbf{0},\lambda_2 \mathbf{v}_i^j)$ and a normal distribution $\bm{\alpha}_{cd} \sim \mathcal{N}(\mathbf{0},\lambda_2\mathbf{v}_i^j\mathbf{I})$.
As shown in Table~\ref{tab:table3}, the segmentation results of these two distribution forms are comparable, validating the robustness of our method to different distributions.

\begin{table}[!t]
\centering
\caption{The influence of different distribution and position schemes for feature augmentation. The best results are highlighted in \textbf{bold}.}
\resizebox{\linewidth}{!}{
    \begin{tabular}{c|ccccc|c}
    \toprule[1pt]
    \multirow{2}*{Method}&Domain1& Domain2& Domain3& Domain4& Domain5& Average\\
    \cline{2-7}
     &DSC $\uparrow$& DSC $\uparrow$& DSC $\uparrow$& DSC $\uparrow$& DSC $\uparrow$& DSC $\uparrow$ \\ \midrule[1pt]
    Normal Distribution &(88.74, 67.56) & (89.75, \textbf{77.76})
& (87.49, 65.67)& (\textbf{81.69}, 67.37)& (\textbf{87.66}, 68.80)& 78.25 \\ 
   Uniform Distribution&({\textbf{88.95}, \textbf{72.55}})& (\textbf{89.86}, 77.61) & (\textbf{88.17}, \textbf{67.22})& (81.09, \textbf{67.50})& (86.67, \textbf{69.19})&\textbf{78.88 }\\ 
    \midrule
    \midrule
    Random position &(87.32, 66.16) & (\textbf{90.19}, {77.00})
& (87.22, 66.85)& (\textbf{81.39}, 67.63)& ({85.86}, 68.56)& 77.82 \\ 
   Maximum $k$ position&({{85.58}, {56.55}})& ({88.76}, 74.71) & (\textbf{88.46}, {65.76})& (80.22, {63.40})& (\textbf{86.96}, {67.86})&{75.83 }\\
    Minimum $k$ position&({\textbf{88.95}, \textbf{72.55}})& ({89.86}, \textbf{77.61}) & ({88.17}, \textbf{67.22})& (81.09, \textbf{67.50})& (86.67, \textbf{69.19})&\textbf{78.88 }\\ 
    \bottomrule[1pt]
    \end{tabular}
    }
\label{tab:table3}
\end{table}

\noindent \textbf{Analysis of perturbation positions.} To verify that perturbations at the positions with the minimum partial derivatives have minimal impact on feature semantics, we conducted a perturbation position analysis experiment across five domains. As shown in Table~\ref{tab:table3}, perturbing the $k$ positions with the minimum partial derivatives resulted in the best performance of the model, confirming the importance of considering direction during the feature moving process.
\section{Conclusion}
We propose a novel domain randomization-based DG method, called ConStyX, for generalizable medical image segmentation. To enable the source domain data to cover a wider range of unseen domains, ConStyX augments the content and style of the training data by moving the deep features toward correct directions with proper degrees. 
Besides, we define three types of augmented features and assign different weights to them to mitigate the negative impact of over-augmented features on model training, while fully leveraging well-augmented features. 
Extensive experiments on five fundus datasets demonstrate that ConStyX achieves compelling performance over the state-of-the-art methods.



%
%
%
 \bibliographystyle{splncs04}
 \bibliography{reference}
%






\end{document}